\documentstyle[prl,aps]{revtex}

\newlength{\defaultparindent}
\newenvironment{Default Paragraph Font}{}{}

\begin{document}
\draft
\title{Luttinger Theorem in One Dimensional Metals}
\author{Krastan B. Blagoev$\thanks{{\it e-mail} {\it blagoev@ph01.bc.edu}}$ and
Kevin S. Bedell$\thanks{{\it e-mail kevin.bedell@bc.edu}}$}
\address{Physics Department, Boston College, Chestnut Hill, MA 02167}
\date{\today }
\maketitle

\begin{abstract}
One dimensional metals are described by Luttinger liquid theory. Recent
experiments have addressed the relation between this non-Fermi liquid
behavior and the existence of a Fermi surface. We show that Luttinger's
theorem, with few modifications, holds for the one-dimensional
Tomonaga-Luttinger model. The implications for the high temperature
superconductors are discussed.
\end{abstract}


Soon after the discovery of the high temperature superconductors ($HTSC$),
Anderson\cite{Anderson87} noted that the two dimensional (2D) copper-oxide
planes in these materials are responsible for the high critical
temperatures. He suggested that the essential properties of the $HTSC$'s are
contained in the physics of the 2D strongly correlated electron liquid.
While there is a consensus on this question, there are debates on the nature
of the low energy physics of that liquid.

The question is whether 2D quantum fluids are described by Landau's Fermi
liquid theory (FLT) or by a theory which resembles the physics of one
dimensional (1D) systems with short range interactions\cite
{Anderson90/91/93,Ren92,Clarke & Strong97}. Perturbation theory shows that
despite some peculiarities compared to three dimensions\cite{Randeria &
Engelbrecht90/91,Metzner & Castellani95}, the essential physics of the 2D
weakly coupled electron liquid is described by Fermi liquid quasiparticles.
However, in 1D perturbation theory is known to violate some of the exact
results and therefore it is feasible that in 2D the same could happen
leading to unknown behavior. 1D metals with short range interactions are
described by Haldane's Luttinger liquid theory\cite{Haldane81} (LLT). In 1D,
the exactly solvable Hubbard\cite{Hubbard63} and Tomonaga-Luttinger\cite
{Tomonaga50,Luttinger63} models capture the essential physics, where instead
of a simple pole, the one-particle Green's function has a branch-cut
singularity and spin and charge propagate with different velocities\cite
{Dzyaloshinskii & Larkin74,Luther & Peschel74} (the Hubbard model in the
metallic phase scales (in renormalization group sense) to the
Tomonaga-Luttinger model\cite{Schultz90,Kawakami & Yang90,Korepin90}). These
properties of 1D systems are also important in understanding the properties
of the organic and inorganic quasi-1D metals. \cite{Smith91,Kim et al.96}.
While we know that in 2D when very strong or long range interactions are
present the Fermi liquid picture breaks and the system can develop charge
or/and spin density waves, or condense into a Wigner crystal, the question
is if this is the case for arbitrary small, short range interactions as is
the case in 1D.

Some time ago, Luttinger\cite{Luttinger60} proved a theorem using
perturbation theory showing that the volume enclosed by the Fermi surface is
an interaction strength invariant (therefore the Fermi sphere is
incompressible) and that the momentum distribution function has a
discontinuity at the Fermi momentum of the noninteracting system\cite
{Abrikosov et al.63}. In 1D, there is no discontinuity in the momentum
distribution function at, $p_F$, the Fermi momentum of the noninteracting
system and the excitations consist only of collective boson modes. The
momentum distribution function, in the vicinity of the Fermi "surface",
behaves as\cite{Solyom79} $|k-p_F|^\alpha $ with nonuniversal, coupling
dependent $\alpha $, and no quasiparticles are present in the liquid. The
absence of quasiparticles near the Fermi surface gives the distinct
properties of the 1D metals. Nevertheless, even in 1D, the Fermi momentum of
the noninteracting system plays an important role when the interactions are
switched on. The exact relationship between Fermi liquid behavior and the
existence of a Fermi surface is not yet clear and we feel it is important to
discuss this relation in Luttinger's theorem. In the case of 1D metals there
have been controversial statements\cite{Haldane94,DiCastro et
al.95,Voit95,Louis95} about the validity of this theorem and this is a
question that we would like to address in this short paper.

In this letter, we prove Luttinger's theorem in 1D and therefore show that
the theorem holds in a system in which the interactions do not produce
quasiparticles: the 1D g-ology model\cite{Solyom79}. This theorem has been
proven order by order in perturbation theory using the general properties of
a Fermi liquid-like Green's function. However, in 1D there exists a closed
integral equation for the single-particle Green's function obtained after
the summation of the perturbation series. A particular case permits an
explicit solution in $k-space$ for the Green's function which is a
double-valued function of the frequency. No $k-space$ solution is available
in the general case, but few results regarding the single-particle Green's
function can be proven without having an explicit expression for its
solutions.

Ignoring umklapp and the backward scattering processes, the g-ology model
describes a set of electrons in 1D with the Hamiltonian\cite{Solyom79}: 
\begin{eqnarray}
H &=&H_0+H_{int} \\
H_0 &=&\sum_{k,\sigma }v_F(k-p_F)a_{k,\sigma }^{\dagger }a_{k,\sigma
}+\sum_{k,\sigma }v_F(-k-p_F)b_{k,\sigma }^{\dagger }b_{k,\sigma } \\
H_{int} &=&\frac 1L\sum_{k_1,k_2,p,\alpha ,\beta }[\Gamma _{\alpha ,\beta
}^2a_{k_1,\alpha }^{\dagger }b_{k_2,\beta }^{\dagger }b_{k_2+p,\beta
}a_{k_1-p,\alpha }  \nonumber \\
&&+\frac 12\Gamma _{\alpha ,\beta }^4(a_{k_1,\alpha }^{\dagger }a_{k_2,\beta
}^{\dagger }a_{k_2+p,\beta }a_{k_1-p,\alpha }  \nonumber \\
&&+b_{k_1,\alpha }^{\dagger }b_{k_2,\beta }^{\dagger }b_{k_2+p,\beta
}b_{k_1-p,\alpha })]
\end{eqnarray}

where $a^{\dagger },a$ ($b^{\dagger },b$) are the creation and annihilation
operators for particles at the $+p_F$ ($-p_F$) branch respectively. $\Gamma
_{\alpha ,\beta }^i=g_{i\parallel }\delta _{\alpha ,\beta }+g_{i\perp
}\delta _{\alpha ,-\beta }$ ($i=2,4$) and $\parallel $ , and $\perp $
correspond to particles with parallel and antiparallel spins respectively.
Using a Ward identity, one obtains Dyson's equation as a singular integral
equation for the single particle Green's function for electrons moving to
the right\cite{Dzyaloshinskii & Larkin74}:

\begin{eqnarray}
G_{+}(p,\epsilon ) &=&G_{0+}(p,\epsilon )  \nonumber  \label{1} \\
&&\times [1+\frac i{4\pi ^2}\int \int dkd\omega G_{+}(p-k,\epsilon -\omega
)K(k,\omega )]\text{,}  \label{2}
\end{eqnarray}
where 
\begin{equation}
K(k,\omega )=\sum\limits_{i=c,s}\left\{ \frac{A_i}{\omega -u_ik+i\delta [k]}+%
\frac{B_i}{\omega +u_ik-i\delta [k]}\right\} \text{,}
\end{equation}
and 
\begin{equation}
G_{0+}(p,\epsilon )=\frac 1{\epsilon -v_F(p-p_F)+i\delta [p-p_F]}
\end{equation}
is the single particle non-interacting Green's function, and the constants $%
A_i$ and $B_i$ depend on the couplings $g_{2\parallel }$ , $g_{2\perp }$
(for particles on different branches) and $g_{4\parallel }$, $g_{4\perp }$
(for particles on the same branch)\cite{Solyom79} (we assume that the
couplings are momentum independent). The charge and spin velocities, $u_c$
and $u_s$ respectively, are functions of the coupling constants. Here $%
p_F=\pi n/2$ and $\delta [q]\equiv \delta sign[q]$.

Luttinger's theorem\cite{Luttinger60,Abrikosov et al.63} states that in $d$
dimensional space: $i$. 
\begin{equation}
2\int_{G(p,0)>0}dp=2\int \theta (p-p_F)dp=V_F=(2\pi )^d\frac NV  \label{5}
\end{equation}
where $p_F$ is the Fermi momentum of the noninteracting system, $N$ is the
mean number of particles in the system, $V$ is the volume of the system, $%
V_F $ is the volume of the Fermi sphere and $G(p,0)$ is the interacting
Green's function. $ii$. the momentum distribution function $n({\bf p})$ has
a discontinuity at the points $\{{\bf p}_F\}$, at which the noninteracting
distribution function $n_0({\bf p})\sim \theta ({\bf p}-{\bf p}_F)$ is
discontinuous. The discontinuity of $n({\bf p})$ is proportional to the
quasiparticle residue, i.e. $\lim\limits_{{\bf p}\rightarrow {\bf p}_F}[n(%
{\bf p}<{\bf p}_F)-n({\bf p}>{\bf p}_F)]=Z$. However, in 1D $Z$ is zero and
the generalized statement is that the derivative of $n({\bf p})$ is singular
with a power law singularity instead of a delta function singularity. We
will show that Luttinger's theorem is satisfied (in 1D) by the Green's
function satisfying Eq.(\ref{2}), i.e. that it changes sign when crossing
the Fermi momentum of the noninteracting system. In this case the
interacting Green's function is a product of (see Eq.(\ref{2})) the
noninteracting Green's function, which changes sign at $p_F$ and a term
which, if it does not change sign when we cross $p_F$, Eq.(\ref{5}) and
therefore the first part of Luttinger's theorem will be satisfied. From now
on we work with particles on the right branch, i.e. moving to the right. Let
us denote by $D(p)$ the term in the parentheses at $\epsilon =0$, i.e. 
\begin{equation}
D(p)=\frac i{4\pi ^2}\int \int dkd\omega G_{+}(p-k,-\omega )K(k,\omega )
\end{equation}
We can integrate (we explain later in the paper in more detail, how these
integrals are calculated) over $\omega $ using the general properties of
fermionic Green's function\cite{Galitskii & Migdal58}: $i$. The
singularities are located in the second quadrant in the complex-frequency
plane for $p-p_F<0$ and in the fourth quadrant for $p-p_F>0$; $ii$. $%
G(p,\omega )\stackrel{\omega \rightarrow \infty }{\rightarrow }\frac 1\omega 
$. The result for $p>p_F$ is 
\begin{eqnarray}
&&\frac 1{2\pi }\sum_{i=c,s}[B_i\int_{-\infty }^0G_{+}(p-k,u_ik)dk  \nonumber
\\
&&-B_i\int_0^\infty G_{+}(p-k,u_ik)\theta (p_F-p+k)dk  \nonumber \\
&&+A_i\int_0^\infty G_{+}(p-k,-u_ik)\theta (p-k-p_F)dk]
\end{eqnarray}
and for $p<p_F$%
\begin{eqnarray}
&&\frac 1{2\pi }\sum_{i=c,s}[B_i\int_{-\infty }^0G_{+}(p-k,u_ik)\theta
(p-k-p_F)dk  \nonumber \\
&&-B_i\int_0^\infty G_{+}(p-k,u_ik)dk  \nonumber \\
&&-A_i\int_{-\infty }^0G_{+}(p-k,-u_ik)\theta (p_F-p+k)dk]
\end{eqnarray}
In the limit $p\rightarrow p_F$ the last two expressions are equal and the
limit is 
\begin{eqnarray}
\lim_{p\rightarrow p_F}D(p) &=&\frac 1{2\pi }\sum_{i=c,s}B_i[\int_{-\infty
}^0G_{+}(p_F-k,u_ik)dk  \nonumber  \label{9} \\
&&-\int_0^\infty G_{+}(p_F-k,u_ik)dk]=-1
\end{eqnarray}
above and bellow $p_F$. Comparison of the exact expression for the momentum
distribution function and its expansion around $p_F$ (both shown later)
shows that 
\begin{equation}
D(p\cong p_F)=-1+const.\mid p-p_F\mid ^\alpha +const.(p-p_F)
\end{equation}
Therefore 
\begin{equation}
G_{+}(p\cong p_F,0)\sim -const.\frac{\mid p-p_F\mid ^\alpha }{p-p_F}
\label{Gf}
\end{equation}
and therefore the Green's function changes sign at $p=p_F$ which completes
the proof of the first part of Luttinger's theorem.

Next, we would like to show that the Green's function given by Eq.(\ref{2})
gives the same number of particles as the noninteracting one. We use a band
width cutoff $A$. Practically, that means that when the Green's function is
integrated over the momentum, one integrates in the interval $[p_F-A,p_F+A]$
and then takes the limit $\Lambda /p_F\rightarrow 0$ where $\Lambda =p_F-A$.
The number of particles will be the same if and only if: 
\begin{eqnarray}
&&\frac 1{4\pi ^2}\lim\limits_{t\rightarrow 0^{+}}\int \int \frac{%
dpd\epsilon }{(2\pi )^2}G_{0+}(p,\epsilon )e^{i\epsilon t}\times  \nonumber
\label{cond} \\
&&\int \int dkd\omega G_{+}(p-k,\epsilon -\omega )K(k,\omega )=0  \label{13}
\end{eqnarray}
since 
\begin{eqnarray}
-2i\lim\limits_{t\rightarrow 0^{+}}\int \int \frac{dpd\epsilon }{(2\pi )^2}%
G_{0+}(p,\epsilon )e^{i\epsilon t} &=&  \nonumber \\
\frac{p_F}\pi \left( 1-\frac \Lambda {p_F}\right) &=&n_{+}
\end{eqnarray}
here $n_{+}$ is the density of particles moving to the right.

Using the mentioned properties of the Green's function, first we integrate
over the frequencies $\epsilon $. Because of the second property, the
integrals on the l.h.s. of Eq.(\ref{13}) are convergent and therefore one
can introduce the limit under the integral and perform it explicitly before
integrating. This allow us to close the contour of integration on either
side of the real frequency axis in the complex plane. Therefore, when the
branch-cut singularity of Green's function and the pole occur on the same
half plane we close the contour in the other half plane and from Cauchy
theorem the corresponding integral is zero. When the pole and the branch-cut
singularity of the Green's function are on the opposite side of the real
axis we choose to close the contour of integration in the half complex plane
where the pole is located and therefore we obtain the branch-cut part
evaluated at that pole. Then we evaluate the second frequency integration in
(\ref{13}) using the same procedure. After performing the double frequency
integration, the $A_i$ terms give zero both in the charge and spin sectors.
This is the reason why the simple square root Green's function\cite
{Dzyaloshinskii & Larkin74} leads to the step-function momentum distribution
function, the same as for the non-interacting case. However, the double
frequency integration of the $B_i$ terms is nonzero and the result is: 
\begin{eqnarray}
n_{+}(p>p_F &&)=\frac 1{2\pi }\sum\limits_{i=c,s}B_i\int
G_{+}(p-k,v_F(p-p_F)+u_ik)\theta (p_F-p+k)\theta (k)dk  \nonumber \\
n_{+}(p<p_F &&)=1+\frac 1{2\pi }\sum\limits_{i=c,s}B_i\int
G_{+}(p-k,v_F(p-p_F)+u_ik)\theta (p-k-p_F)\theta (-k)dk
\end{eqnarray}
The exact value of the last two integrals cannot be obtained without an
explicit expression for the function. However, from the exact solution\cite
{Mattis and Lieb65} and from perturbative calculations\cite{Dzyaloshinskii &
Larkin74,Solyom79,Voit93}, it is known that the momentum distribution
function is continuous and without a jump at $p=p_F$ and for $p\simeq p_F$
is 
\begin{eqnarray}
n(p &>&p_F)\sim \frac 12-C_1\mid p-p_F\mid ^\alpha -C_2(p-p_F)  \nonumber \\
n(p &<&p_F)\sim \frac 12+C_1\mid p-p_F\mid ^\alpha -C_2(p-p_F)
\end{eqnarray}
where $C_1$ and $C_2$ are constants\cite{Voit93}. From the last two
representations of the momentum distribution function follows the behavior
of the integrals of the type encountered in the expression for $G_{+}(p,0)$
in the neighborhood of $p_F$.

In the perturbation theory accessible regime ($\alpha <1$) the derivative $%
\frac{dn}{dp}$ $\sim |p-p_F|^{\alpha -1}$ is singular approaching infinity
with a power law. We adopt the general definition that the Fermi surface is
the set of $k$-points at which the $mth$ derivative of the momentum
distribution function has a singularity i.e., $\{p_F\}\equiv \{\forall k:%
\frac{d^mn(k)}{dk^m}$ is singular$\}$. These we shall call Fermi points of
order $m$. In the usual Fermi liquid the Fermi surface consist of zero order
Fermi points while in the Luttinger liquid these are of first order.
Although the zero order Fermi surface has disappeared, the first order Fermi
surface is left and the generalized statement of the second part of
Luttinger's theorem holds. When $\alpha >1$ liquid droplets form and as long
as $\alpha $ is not an integer the derivative is zero at $p_F$, but there
will exist a number $m$ so that $\frac{d^mn}{dp^m}$ is singular at $p_{F%
\text{ }}$and this will correspond to a Fermi surface of order $m$.

At the end we must show that the following expression is zero in the limit $%
\Lambda \rightarrow 0$: 
\begin{eqnarray}
&&I=\frac 1{2\pi }\sum_{i=c,s}B_i[\int\limits_\Lambda
^{p_{_F}}dp\int\limits_{-\infty }^{p-p_{_F}}dqG_{+}(q,v_F(p-p_F)+u_i(p-q)) 
\nonumber \\
&&+\int\limits_{p_F}^{2p_{_F}-\Lambda }dp\int\limits_{p-p_{_F}}^\infty
dqG_{+}(q,v_F(p-p_F)+u_i(p-q))]  \label{19}
\end{eqnarray}
Taking into account the two representations of the momentum distribution
function one sees that in the above mentioned limit the two integrals cancel
each other and the total number of particles on the right branch is $%
n_{+}=p_F/\pi $.

In this paper, we have shown from Dyson's equation, that Luttinger's theorem
holds for the one-dimensional Tomonaga-Luttinger model. In general, the
theorem is based on the counting of the fermionic degrees of freedom before
and after the interactions are turned on. In the case of Fermi liquids, the
one-to-one correspondence between the noninteracting particles and the
quasiparticles ensures the validity of the theorem. In the case of 1D
Luttinger liquids, the number of charge particles in the interacting system
is exactly equal to the number of electrons in the noninteracting system.
Our conjecture is that as long as the number of states and excitations with
and without interactions are the same, Luttinger's theorem will be
satisfied. Some recently proposed 2D Luttinger-type Green's functions
satisfy this condition\cite{Yin & Chakravarty96,Ren92}. The $HTSC$'s, at
zero doping and bellow the N\'{e}el temperature are antiferromagnetic
insulators. Doping destroys the antiferromagnetic order and a metallic phase
occurs above the superconducting critical temperature. Our discussion
indicates that in the high-temperature superconducting cuprates, Luttinger's
theorem will be satisfied in this temperature and doping interval regardless
of the nature of the electronic liquid (Fermi or Luttinger).

We would like to thank J. Voit for the detailed analysis of this paper and
the helpful suggestions. Special thanks are also due to J. Engelbrecht and
M. Gul\'{a}csi for the insightful suggestions and to A. Nazarenko for
carefully reading the paper. We would like to thank P.W. Anderson, C.
DeLeone, I.E. Dzyaloshinskii, and G. Zimanyi, for the useful discussions of
this topic.

\end{document}